\documentclass[longauth]{aa}  
\usepackage{rotating}
\usepackage{amsmath}
\usepackage[colorlinks=true,citecolor=blue, urlcolor=blue]{hyperref}

\usepackage{graphicx}
\usepackage{txfonts}
\usepackage{amssymb}
\usepackage{lscape}
\usepackage[bf,nooneline]{subfigure}
\bibpunct{(}{)}{;}{a}{}{,}

\begin{document}

\title{The VIMOS Public Extragalactic Redshift Survey (VIPERS)}
\subtitle{Environment-size relation of massive passive galaxies at $0.5 \leqslant z \leqslant 0.8$ 
\thanks{based on observations collected at the European Southern Observatory, Cerro Paranal, Chile, using the Very Large Telescope under programs 182.A-0886 and partly 070.A-9007. 
Also based on observations obtained with MegaPrime/MegaCam, a joint project of CFHT and CEA/DAPNIA, at the Canada-France-Hawaii Telescope (CFHT), which is operated by the
National Research Council (NRC) of Canada, the Institut National des Sciences de l’Univers of the Centre National de la Recherche Scientifique (CNRS) of France, and the University of Hawaii. This work is based in part on data products produced at TERAPIX and the Canadian Astronomy Data Centre as part of the Canada-France-Hawaii Telescope Legacy Survey, a collaborative project of NRC and CNRS. The VIPERS web site is http://www.vipers.inaf.it/. }}
\author{
A. Gargiulo\inst{\ref{iasf-mi}}
\and O.~Cucciati\inst{\ref{oabo}}
\and B.~Garilli\inst{\ref{iasf-mi}} 
\and M.~Scodeggio\inst{\ref{iasf-mi}}
\and M.~Bolzonella\inst{\ref{oabo}} 
\and G.~Zamorani\inst{\ref{oabo}}
\and G.~De Lucia\inst{\ref{oats}}
\and J.~Krywult\inst{\ref{kielce}}
\and L.~Guzzo\inst{\ref{brera},\ref{unimi}}              
\and B.~R.~Granett\inst{\ref{brera},\ref{unimi}}
\and S.~de la Torre\inst{\ref{lam}}       
\and U.~Abbas\inst{\ref{oa-to}}
\and C.~Adami\inst{\ref{lam}}
\and S.~Arnouts\inst{\ref{lam}} 
\and D.~Bottini\inst{\ref{iasf-mi}}
\and A.~Cappi\inst{\ref{oabo},\ref{nice}}           
\and P.~Franzetti\inst{\ref{iasf-mi}}   
\and A.~Fritz\inst{\ref{iasf-mi}}      
\and C.~Haines\inst{\ref{cile}}
\and A.~J.~Hawken\inst{\ref{brera},\ref{unimi}} 
\and A.~Iovino\inst{\ref{brera}}
\and V.~Le Brun\inst{\ref{lam}}
\and O.~Le F\`evre\inst{\ref{lam}}
\and D.~Maccagni\inst{\ref{iasf-mi}}
\and K.~Ma{\l}ek\inst{\ref{warsaw-nucl}}
\and F.~Marulli\inst{\ref{unibo},\ref{infn-bo},\ref{oabo}} 
\and T.~Moutard\inst{\ref{halifax},\ref{lam}}  
\and M.~Polletta\inst{\ref{iasf-mi},\ref{marseille-uni},\ref{toulouse}}
\and A.~Pollo\inst{\ref{warsaw-nucl},\ref{krakow}}
\and L.A.M.~Tasca\inst{\ref{lam}}
\and R.~Tojeiro\inst{\ref{st-andrews}}  
\and D.~Vergani\inst{\ref{oabo}}
\and A.~Zanichelli\inst{\ref{ira-bo}}
\and J.~Bel\inst{\ref{cpt}}
\and E.~Branchini\inst{\ref{roma3},\ref{infn-roma3},\ref{oa-roma}}
\and J.~Coupon\inst{\ref{geneva}}
\and O.~Ilbert\inst{\ref{lam}}
\and L.~Moscardini\inst{\ref{unibo},\ref{infn-bo},\ref{oabo}}
\and J.A.~Peacock \inst{\ref{edi}}
}
\institute{
INAF - Istituto di Astrofisica Spaziale e Fisica Cosmica Milano, via Corti 12, 20133 Milano, Italy \label{iasf-mi}
\and INAF - Osservatorio di Astrofisica e Scienza dello Spazio di
Bologna, via Piero Gobetti 93/3, I-40129 Bologna, Italy, Italy \label{oabo} 
\and Institute of Physics, Jan Kochanowski University, ul. Swietokrzyska 15, 25-406 Kielce, Poland \label{kielce}
\and INAF - Osservatorio Astronomico di Trieste, via G. B. Tiepolo 11, 34143 Trieste, Italy \label{oats}
\and INAF - Osservatorio Astronomico di Brera, Via Brera 28, 20122 Milano --  via E. Bianchi 46, 23807 Merate, Italy \label{brera} 
\and  Universit\`{a} degli Studi di Milano, via G. Celoria 16, 20133 Milano, Italy \label{unimi}
\and Aix Marseille Univ, CNRS, LAM, Laboratoire d'Astrophysique de Marseille, Marseille, France  \label{lam}
\and INAF - Osservatorio Astrofisico di Torino, 10025 Pino Torinese, Italy \label{oa-to}
\and Laboratoire Lagrange, UMR7293, Universit\'e de Nice Sophia Antipolis, CNRS, Observatoire de la C\^ote d'Azur, 06300 Nice, France \label{nice}
\and Instituto de Astronom\'{i}a y Ciencias Planetarias, Universidad de Atacama, Avenida Copayapu 485, Copiap\`{o}, Chile \label{cile}
\and Dipartimento di Fisica e Astronomia - Alma Mater Studiorum Universit\`{a} di Bologna, via Piero Gobetti 93/2, I-40129 Bologna, Italy \label{unibo}
\and National Centre for Nuclear Research, ul. Pasteura 7, 02-093 Warszawa, Poland \label{warsaw-nucl}
\and INFN, Sezione di Bologna, viale Berti Pichat 6/2, I-40127 Bologna, Italy \label{infn-bo}
\and Department of Astronomy $\&$ Physics, Saint Mary's University, 923 Robie Street, Halifax, Nova Scotia, B3H 3C3, Canada \label{halifax}
\and Aix-Marseille Universita, Jardin du Pharo, 58 bd Charles Livon, F-13284 Marseille cedex 7, France \label{marseille-uni}
\and IRAP,  9 av. du colonel Roche, BP 44346, F-31028 Toulouse cedex 4, France \label{toulouse} 
\and Astronomical Observatory of the Jagiellonian University, Orla 171, 30-001 Cracow, Poland \label{krakow} 
\and School of Physics and Astronomy, University of St Andrews, St Andrews KY16 9SS, UK \label{st-andrews}
\and INAF - Istituto di Astrofisica Spaziale e Fisica Cosmica Bologna, via Gobetti 101, I-40129 Bologna, Italy \label{iasf-bo}
\and INAF - Istituto di Radioastronomia, via Gobetti 101, I-40129, Bologna, Italy \label{ira-bo}
\and Canada-France-Hawaii Telescope, 65--1238 Mamalahoa Highway, Kamuela, HI 96743, USA \label{cfht}
\and Aix Marseille Univ, Univ Toulon, CNRS, CPT, Marseille, France \label{cpt}
\and Dipartimento di Matematica e Fisica, Universit\`{a} degli Studi Roma Tre, via della Vasca Navale 84, 00146 Roma, Italy\label{roma3} 
\and INFN, Sezione di Roma Tre, via della Vasca Navale 84, I-00146 Roma, Italy \label{infn-roma3}
\and INAF - Osservatorio Astronomico di Roma, via Frascati 33, I-00040 Monte Porzio Catone (RM), Italy \label{oa-roma}
\and Department of Astronomy, University of Geneva, ch. d'Ecogia 16, 1290 Versoix, Switzerland \label{geneva}
\and Institute for Astronomy, University of Edinburgh, Royal  Observatory, Blackford Hill, Edinburgh EH9 3HJ, UK \label{edi}
 }
   \date{Received 12 June 2018; accepted **}
  \abstract
{We use the unparalleled statistics of the VIPERS survey to investigate the relation between the surface mean stellar mass density $\Sigma = \cal{M}$/$(2\,\pi\it{R_e}^{2})$ of massive passive galaxies (MPGs, $\cal{M} \geqslant$ 10$^{11}$M$_{\odot}$) and their local environment in the redshift range $ 0.5 \leqslant z \leqslant 0.8$. 
Passive galaxies were selected on the basis of their NUV$r$K colors ($\sim$900 objects), and the environment was defined as the galaxy density contrast, $\delta$, using the fifth$^{}$ nearest-neighbor approach. The analysis of $\Sigma \text{ versus }\delta$ was carried out in two stellar mass bins. In galaxies with $\cal{M} \leqslant$2$\times$10$^{11}$M$_{\odot}$, no correlation between $\Sigma$ and $\delta$ is observed. This implies that the accretion of satellite galaxies, which is more frequent in denser environments (groups or cluster outskirts) and efficient in reducing the galaxy $\Sigma$, is not relevant in the formation and evolution of these systems. 
Conversely, in galaxies with $\cal{M} >$2$\times$10$^{11}$M$_{\odot}$, we find an excess of MPGs with low $\Sigma$ and a deficit of high-$\Sigma$ MPGs in the densest regions with respect to other environments. We interpret this result as due to the migration of some high-$\Sigma$ MPGs ($<1\%$ of the total population of MPGs) into low-$\Sigma$ MPGs, probably through mergers or cannibalism of small satellites. In summary, our results imply that the accretion of satellite galaxies has a marginal role in the mass-assembly history of most MPGs. 

We have previously found that the number
density of VIPERS massive star-forming galaxies (MSFGs) declines rapidily from $z$\,=\,0.8 to $z$\,=\,0.5, which mirrors the rapid increase in the number density of MPGs. This indicates that the MSFGs at $z \geqslant 0.8$ migrate to the MPG population.
Here, we investigate the $\Sigma-\delta$ relation of MSFGs at $z \geqslant 0.8$ and find that it is consistent within 1$\sigma$ with that of low-$\Sigma$ MPGs at $z \leqslant 0.8$.  Thus, the results of this and our previous paper show that MSFGs at $z \geqslant 0.8$ are consistent in terms of number and environment with being the progenitors of low-$\Sigma$ MPGs at $z < 0.8$. 
}
   \keywords{galaxies: elliptical and lenticular, cD; galaxies: formation; galaxies: evolution; 
              galaxies: high redshift}
\titlerunning {}
   \authorrunning {Gargiulo et al.}

   \maketitle
%
\section{Introduction}\label{introduction}

It is now well established that the physical parameters of galaxies (e.g., star formation timescale, formation redshift, and morphology) are closely correlated with the environment that these galaxies reside in and with their stellar mass  \citep[e.g.,][]{dressler80,hashimoto98,gavazzi06,kauffmann04,saracco17,tamburri14,cucciati17,balogh04,cucciati06,elbaz07,bundy06,davidzon16,peng10,thomas10,pozzetti10,haines07}. These correlations indicate that the stellar mass-assembly history in a galaxy depends both on stellar mass and environment \citep[e.g.,][]{peng10}. Because these two parameters are closely related and depend on redshift, understanding the main processes that are responsible for galaxy formation and evolution requires a careful multi-parameter investigation.

How passive galaxies (PGs) form and evolve is still unclear. It has been shown that at fixed stellar mass,  the mean effective radius $\langle {\it R}_e \rangle$ (where ${\it R}_e$ is the radius that encloses half of the total light) of PGs increases by a factor 4 between $z = 2$ and $z = 0$ \citep[e.g.,][]{vanderwel14}.
To explain the observed size evolution, a scenario involving galaxy mergers has been proposed because the continuous accretion of satellite galaxies increases both mass and radius of a galaxy. 
In this scenario, a highly dissipative process (e.g., a gas-rich merger, Hopkins et al. \citeyear{hopkins08}{}) at $z \gtrsim 2$ forms the compact passive cores we observe at high-$z$. In the subsequent 10\,Gyr, these compact galaxies grow a low-density halo through the dry accretion of small satellites (e.g., through dry minor mergers, cannibalism) until they match the typical size of local PGs \citep[e.g.,][]{robertson06,naab07, naab09, hilz13, vandokkum08, vandokkum10}. Dry mergers are expected to be more frequent in groups than in the field~\citep[e.g.,][]{treu03}, and cannibalism is more recurrent for central galaxies, therefore this model predicts a positive correlation between environmental density and galaxy size.

Several studies have found a positive correlation between galaxy size and environment at  $0.7 \lesssim z \lesssim 2$, with cluster galaxies 30-50$\%$ larger than field counterparts of similar stellar mass \citep[e.g.,][]{cooper12,papovich12,bassett13,strazzullo13,lani13,delaye14}. Conversely, no correlation is found in the local Universe between the mean galaxy size and the environment \citep[][but see also \citealt{poggianti13, cebrian14}]{maltby10,nair10,huertascompany13b,cappellari13env}. This apparent discrepancy between the dependency of size on environment at high- and low-$z$ can be reconciled within a scenario of hierarchical assembly. In this framework, cluster galaxies evolve faster (i.e., become larger earlier) than galaxies in the field because more mergers per unit time occur in high-density regions \citep[e.g.][]{delucia04, andreon18}.

However, an increasing number of independent studies  have cast doubts on the relevance of dry accretion of satellites in the assembly history of passive galaxies and suggested instead that they are mostly star-forming systems that have progressively halted their star formation \citep[e.g.,][]{lilly16, haines17}. \citet{gargiulo17} (hereafter Paper I) found that the increase in number density of massive passive galaxies (MPGs, ${\cal M}$ $\geqslant$ 10$^{11}$M$_{\odot}$) from $z$ $\sim$ 0.8 to $z \sim 0.5$ is due to the continuous addition of large galaxies with low surface stellar mass density $\Sigma$ ($\Sigma$ = M$_{\star}$/(2$\pi\it{R_e}^{2}$) $\leqslant$ 1000 M$_{\odot}$pc$^{-2}$). This increase is entirely accounted for by the observed decrease in the number density of  massive star-forming galaxies (MSFGs, i.e., all non-passive $\cal{M} \geqslant$ 10$^{11}$M$_{\odot}$ objects) in the same redshift range. This evidence indicates that these star-forming systems could represent the progenitors of the subsequent emerging class of larger MPGs. In this scenario no correlation between galaxy size and environment is expected at any redshift. Other studies at high redshift find that the galaxy size does not depend on the environment \citep[e.g.,][]{huertascompany13,damjanov15env,kelkar15,saracco14,saracco17,rettura10,newman14,allen15}. The lack of size-environment trends at low- and high-$z$, coupled with the results on the number densities,
contradicts the expectations of galaxy evolutionary models that are based on hierarchial assembly of stellar matter. 
The study of the mean size  (or of the mean surface stellar mass density) of PGs as a function of environment is a powerful probe for models of galaxy formation and evolution. In particular, the recent findings described above clearly show that the size-environment relation at high-$z$ for passive galaxies needs to be better defined.

In  this  context, MPGs deserve  particular  attention.  These  systems  are  expected  to evolve mainly through (dry) mergers \citep[e.g.,][]{hopkins09, delucia07}. If this is the case, we should detect a stronger signal in this mass range for the size - environment relation with respect to a lower mass range. Because MPGs are extremely rare, very few studies so far have investigated the size-environment relation for this class of objects at high-$z$

We here take advantage of the unparalleled large statistics of the VIMOS Public Extragalactic Redshift Survey (VIPERS, e.g., Guzzo et al.  \citeyear{guzzo14}, Scodeggio et al. \citeyear{scodeggio16}) to investigate the $\Sigma$-environment relation for MPGs at $0.5 \leqslant z \leqslant 0.8$.  In Section 2 we describe the MPG sample selection and define the environment in terms of the density contrast $\delta$. In Section 3 we determine the $\Sigma$-$\delta$ relation and analyze it as a function of stellar mass. Finally, in Section 4 we compare the $\Sigma$-$\delta$ relation found for MPGs with that of their active counterparts at higher $z$. Our results and conclusions are summarized in Section 5.
Throughout the paper we adopt the Chabrier (2003) initial mass function (IMF) and a flat $\Lambda$CDM cosmology with $\Omega_M = 0.3$, $\Omega_\Lambda = 0.7$ and H$_{0}$ = 70\,km\,s$^{-1}$\,Mpc$^{-1}$. Effective radii are circularized, meaning that $\it{R_e} = \sqrt{ab}$, where $a$ and $b$ are the semi-major and semi-minor axes, respectively, of the isophote that contains half of the total light.

\section{Data}
\label{data}
The data used in this paper are taken from a beta version of the final release of the VIPERS survey \citep{scodeggio16}\footnote{www.vipers.inaf.it}. The data set used here is almost identical to the publicly released PDR-2 catalog, with the exception of a subset of a few redshifts (mostly at $z > 1.2$, thus beyond our range of investigation). In order to be consistent with Paper I, we do not use the final release.

VIPERS is an ESO large program that has measured redshifts for 89,128 galaxies to $i_{AB}=22.5$, distributed over an effective area of 16.3 deg$^{2}$. Spectroscopic targets were selected in the W1 and W4 fields of the Canada-France-Hawaii Telescope (CFHT) Legacy Survey Wide. Using the multiband catalog of the CFHTLS Wide  photometric survey (release T0005, completed by the subsequent T0006), we selected as targets all the objects that satisfied the following two conditions:
\begin{eqnarray}
i_{AB} \leqslant 22.5, \\
(r - i) > 0.5\times(u - g)\,\,\, OR \,\,\,\, (r - i) > 0.7
,\end{eqnarray}
where i$_{AB}$ is corrected for extinction. The relatively bright magnitude limit ensures useful spectral quality in a limited exposure time, while the color-color preselection in the \textit{ugri} plane minimizes the number of galaxies at $z \lesssim$ 0.5 \citep[for details, see][]{garilli14}. No additional criteria using photometric redshift were imposed. The observations were carried out with VIMOS at the Very Large Telescope (VLT) using the low-resolution red grism (R $\sim$ 220), which covers the wavelength range 5500 - 9500\,\smash{\AA}. Within each of the four VIMOS quadrants, slits were  on average assigned to $\sim$47$\%$ of the possible targets, thus defining the  target  sampling  rate (TSR). The {\it rms} error of the measured redshifts has been estimated to be $\sigma_{z}(z)$ = 0.00054$\times$(1+$z$) \citep{scodeggio16}.
A complete description of the PDR-2 data release can be found in Scodeggio et al. (2018) and references therein. 

Stellar masses and absolute magnitudes were derived by fitting the $ugriz$K photometry with a grid of composite stellar population models \citep{bruzual03} for the whole VIPERS sample. The models assume an exponentially declining star formation history $e^{-t/\tau}$ ($\tau$ = [0.1-30]Gyr), and solar and subsolar (0.2Z$_{\odot}$) metallicities. Dust attenuation was modeled assuming the \citet{calzetti00} and the \citet{prevot84} prescriptions (see more details in Davidzon et al. \citeyear{davidzon13,davidzon16} for the fit and in Moutard et al. \citeyear{moutard16} for the photometry). 

Structural parameters (e.g., $R_e$, S\'ersic index $n$) for the whole VIPERS sample were derived from the $i$-band CHFTLS-Wide images by fitting the light profile of the galaxies with a 2D point spread function (PSF) convolved S\'ersic profile \citep{krywult16}. The fit was performed with the GALFIT code \citep{peng02}. The CFHTLS images have a pixel scale of $\sim$0.187$''$/px, and the full width at half-maximum (FWHM) of point-like sources varies from $\sim$0.5$''$ to 0.8$''$. The PSF is successfully modeled  with a 2D Chebychev approximation of the elliptical Moffat function over $\sim$ 90$\%$ of the whole VIPERS area. In regions that lack suitable bright and unsaturated stars for determining the local PSF or at the edge of the images, it was not possible to reconstruct a reliable PSF model. Because these regions are well defined, we removed them from our morphological analysis. 
Errors of the fitting parameters were derived through simulations. In particular, the error on $\it{R_e}$ is lower than 4.4$\%$ (12$\%$) for 68\% (95\%) of the sample.

\subsection{Galaxy environment}
\label{data1}

The environment in the VIPERS fields is characterized by the galaxy density contrast $\delta$, which is described in~\citet[][see also \citeauthor{davidzon16} 2016]{cucciati14, cucciati17} and is defined as
\begin{equation}
\delta(RA,DEC,z) = [\rho(RA,DEC,z) - \bar{\rho(z)}]/\bar{\rho(z)},
\end{equation}
where $\rho(RA,DEC,z)$ is the local number density of the tracers (see below) within a certain volume (or filter window) centered at the galaxy position (${\it RA,DEC,z}$), and $\bar{\rho(z)}$ is the mean number density at that redshift.

In this analysis we refer to the local overdensity $\delta$ that is derived
by adopting as filter window a cylinder with center on the galaxy. Its depth is
$\pm1000$ km/s, and the radius is equal to the distance of the fifth nearest
tracer.

Because VIPERS is a flux-limited survey, it includes galaxies with a lower luminosity limit that becomes increasingly brighter with redshift. To compute the density contrast using a homogeneous
galaxy population in the entire redshift range available for VIPERS, we used a
“volume-limited” sample (our ``tracers'') that comprises galaxies with both
spectroscopic and photometric redshift. This sample was selected to
satisfy M$_{B} \leqslant$ ($-20.4 - z$), and based on the VIPERS flux limit, it is complete for $z \leqslant 0.9$.  The redshift dependence of the luminosity thresholds is designed to account for evolutionary effects because it roughly follows the same dependence on redshift as the characteristic luminosity of the galaxy luminosity function \citep[see, e.g.,][]{kovac10}. \citet{cucciati17} also estimated the density contrast field with a sample of tracers limited to M$_{B} \leqslant$ ($-20.9 - z$). This brighter sample is complete up to
$z = 1$, but it is sparser than the sample with M$_{B} \leqslant$ ($-20.4 - z$), which means that the local density is computed on average on larger scales, that is, with a lower resolution.

For this analysis, we chose the density contrast field that was computed
using the tracers with M$_{B} \leqslant$ ($-20.4 - z$) because it is a good compromise between the allowed redshift extension ($z \leqslant 0.9$) and the scale down to which the field is reconstructed (a factor$\sim2$
smaller than with the brighter sample).
In order to study the effect of the environment on the mass assembly of MPGs, we subdivided the sample of MPGs into four subsamples according to the quartiles of the 1+$\delta$ distribution (see Sec. 3). To take the possible variation of the 1+$\delta$ distribution with $z$ into consideration, we estimated the quartile values in four redshift bins of equal width 0.1. In Appendix A we show that the 1+$\delta$ distribution of MPGs at $0.8 < z \leqslant 0.9$ is significantly different from the 1+$\delta$ distribution at $0.5 \leqslant z \leqslant 0.8$. For this reason we limit our analysis at $z \leqslant 0.8$.

\subsection{Final sample}
\label{data2}

Following Paper I, we identified the VIPERS MPGs by selecting from the VIPERS spectroscopic catalog all of the galaxies  $i$) with secure redshift measurement (confidence level $>$ 95$\%$, i.e., quality flag 2 $\leqslant zflag \leqslant$ 9.5), $ii$) in the redshift range 0.5 $\leqslant z \leqslant$ 0.8, $iii$) with $\cal{M} \geq$ 10$^{11}$M$_{\odot}$ , and $iv$) defined as passive on the basis of their rest-frame near-ultraviolet (NUV) -$r$ and $r$-K colors. As in \citet{davidzon16} and similarly to Paper I, we defined as quiescent those galaxies for which
\begin{eqnarray} 
{\rm NUV} - {\rm r} &>& 3.75, \\ 
 {\rm NUV} - {\rm r} &>& 1.37 \times ({\rm r} - {\rm K}) + 3.2, \\ 
 {\rm r} - {\rm K} &<& 1.3.
\end{eqnarray}
As shown in \citet{davidzon16} (see also Davidzon et al. \citeyear{davidzon13}), VIPERS is 100$\%$ complete at $z = 0.8$  both for passive and star-forming galaxies with $\cal{M} \geq$ 10$^{11}$M$_{\odot}$.

\begin{figure}
\includegraphics[width=9.5cm]{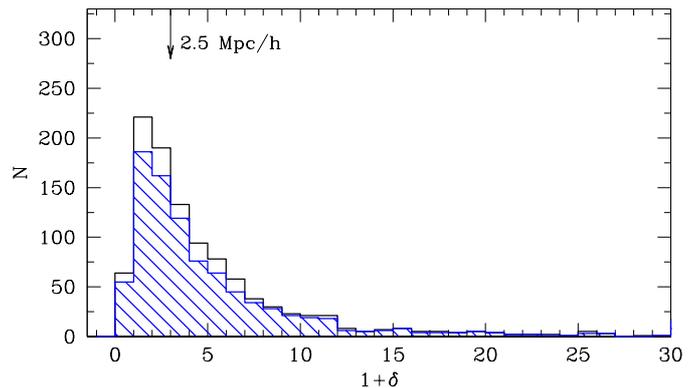}
        \caption{Distribution of the galaxy density contrast for MPGs at 0.5 $\leqslant z \leqslant$ 0.8 with reliable $\delta$ estimates (black solid line). The arrow at the top of the plot is the median value of the distribution with the corresponding value of the projected distance of the fifth nearest neighbor in units of Mpc/h. The solid blue dashed histogram shows the distribution of 1+$\delta$ for the MPGs at 0.5 $\leqslant z \leqslant$ 0.8 with reliable $\delta$ and $R_{e}$ estimates.}
    \label{deltadist}
\end{figure}

When the local density $\delta$ is computed, the projected cylinder can fall outside the area of the survey for galaxies near the edges of the survey or near the CCD gaps. This alters the local density estimate. \citet{cucciati17} showed  that an accurate reconstruction of the local density is provided when more than 60$\%$ of the volume of the cylinder is within the survey area \citep[for more details, see][]{cucciati17, davidzon16}. For this reason,  we included  only the galaxies in our sample that satisfied this condition. In Fig.\ref{deltadist} we report the 1+$\delta$ distribution of the selected sample (black line), its median value (black arrow), and the corresponding projected distance of the fifth nearest neighbor in Mpc/h unit.

Similarly to the $\delta$ estimate, a reliable $R_{e}$ is not available for a small portion of VIPERS galaxies  even when the PSF is well modeled. This is either because the algorithm does not converge or because the best-fit values of $n$ $<$ 0.2 are unphysical \citep[see][]{krywult16}. These objects were removed from the sample. This choice reduced the sample to 902 MPGs. This is the final sample we used for the MPG $\Sigma$-$\delta$ analysis.

In Fig. \ref{deltadist} the blue histogram shows the local density distribution of the final sample. We verified that the local density distributions of galaxies with and without a reliable $\it{R_e}$ measurement are consistent with each other (the Kolmogorov-Smirnov probability that the two distributions shown in Fig. \ref{deltadist} are extracted from the same parent population is 0.99). 
At the same time, we verified that the lack of a reliable $\delta$ estimate does not depend on $\Sigma$.

For each MPG of the final sample we derived the surface mean stellar mass density $\Sigma.$  Consistently with Paper I, we defined a low-$\Sigma$ sample, an intermediate-$\Sigma$ sample, and a high-$\Sigma$ sample composed of MPGs with $\Sigma \leqslant$ 1000\,M$_{\sun}$\,pc$^{-2}$, $1000 < \Sigma \leqslant$ 2000\,M$_{\sun}$\,pc$^{-2}$ , and $\Sigma >$ 2000\,M$_{\sun}$\,pc$^{-2}$, respectively. 

\begin{figure}
\includegraphics[width=9.5cm]{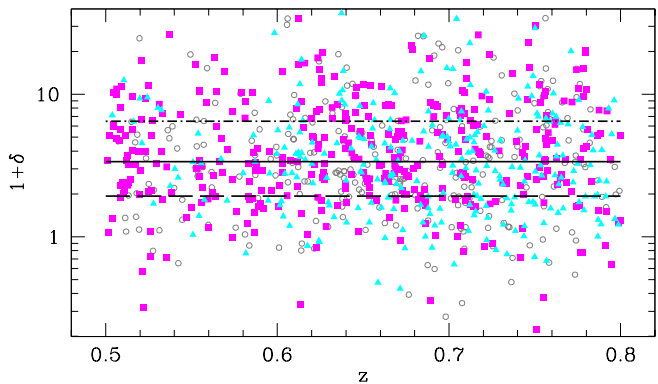}
        \caption{Galaxy density contrast as a function of $z$ for the population of MPGs with reliable estimates  of $\delta$ and $\it{R_e}$  (gray circles, magenta squares, or cyan triangles). Solid lines indicate the median value of the $(1 + \delta)$ distribution, while the 25th and the 75th percentiles are indicated with long-dashed and dot-dashed lines, respectively. Magenta filled squares are MPGs with $\Sigma \leqslant$ 1000\,M$_{\sun}$\,pc$^{-2}$ (i.e., the low-$\Sigma$ sample), cyan filled triangles are MPGs with $\Sigma >$ 2000\,M$_{\sun}$\,pc$^{-2}$ (i.e., the high-$\Sigma$ sample) and gray circles are MPGs with  1000 $< \Sigma \leqslant$ 2000\,M$_{\sun}$\,pc$^{-2}$. }
    \label{delta}
\end{figure}

We repeat that to study the effect of the environment on the size of the MPGs, we subdivided the sample of MPGs into four subsamples according to the quartiles of the 1+$\delta$ distribution. 
In Fig. \ref{delta} we show the local density $\delta$ of MPGs as a function of $z$ and the 25th, 50th, and 75th percentiles of the 1+$\delta$ distribution at $0.5 \leqslant z \leqslant 0.8$ (see Table \ref{sample} for the percentile values). 
Hereafter, we refer to the four density bins as D1 ($(1 + \delta) <$  25th percentile value), D2 (25th $\leqslant (1 + \delta) <$ 50th percentile value), D3 (50th $\leqslant (1 + \delta) <$ 75th percentile value), and D4 ($(1 + \delta) \geqslant$  75th percentile value). Most galaxies in the highest density bin (i.e., $(1 + \delta) >$ 6.48) are in (small) groups, as shown by \citet{cucciati17} (see their Fig. 2). In the same paper, the authors showed that more than 50$\%$ of the galaxies are expected to be central galaxies  at our $\mathcal{M}$ scales and redshift range, (see Fig. 8 in \citeauthor{cucciati17} 2018). This implies that moving from the D1 to D4 density bin, the probability increases that a galaxy assembles stellar mass through the accretion of satellites (through a dry merger or cannibalism).

 \begin{table*}
 \caption{Number of MPGs and 1+$\delta$ percentiles in the full MPG sample and in subsamples subdived by $z$ and $\mathcal{M}$.}
   \begin{center}
   \begin{tabular}{cccccccccc}
   \hline
   \hline
     & $z$ & $N_{tot}$ & $N_{D1}$ & $N_{D2}$ & $N_{D3}$ & $N_{D4}$ & 25$^{th}$ & 50$^{th}$ & 75$^{th}$ \\
 \hline
\multicolumn{10}{c}{Total MPG sample}\\
 \hline
       All         & 0.5 $\leq z \leq$ 0.8 & 902 & 226 & 225  & 226 & 225 & 1.94 & 3.37 & 6.48 \\
$\cal{M}$ $\leq$ 2$\times$10$^{11}$M$_{\odot}$ & 0.5 $\leq z \leq$ 0.8 & 742 & 186 & 186  & 185 & 185 & 1.90 & 3.26 & 6.18 \\
$\cal{M}$ $>$ 2$\times$10$^{11}$M$_{\odot}$ & 0.5 $\leq z \leq$ 0.8 & 160 & 40 & 40  & 40 & 40 & 2.22 & 3.94 & 6.81 \\
                   
\hline
\multicolumn{10}{c}{Low-$\Sigma$ MPG sample}\\
 \hline
             All                               & 0.5 $\leq z \leq$ 0.8 & 386 & 80 & 102 & 95 & 109 & 1.94 & 3.37 & 6.48 \\  
$\cal{M}$ $\leq$ 2$\times$10$^{11}$M$_{\odot}$ & 0.5 $\leq z \leq$ 0.8 & 313 & 67 & 84  & 82 & 80 & 1.90 & 3.26 & 6.18 \\
$\cal{M}$ $>$ 2$\times$10$^{11}$M$_{\odot}$ & 0.5 $\leq z \leq$ 0.8 & 73 & 14 & 17  & 17 & 25 & 2.22 & 3.94 & 6.81  \\
 \hline
\multicolumn{10}{c}{High-$\Sigma$ MPG sample}\\ 
  \hline
           All                                & 0.5 $\leq z \leq$ 0.8 & 255  & 78 & 56 & 65 & 56 & 1.94 & 3.37 & 6.48 \\
$\cal{M}$ $\leq$ 2$\times$10$^{11}$M$_{\odot}$ & 0.5 $\leq z \leq$ 0.8 & 221 & 65 & 48  & 52 & 56 & 1.90 & 3.26 & 6.18 \\
$\cal{M}$ $>$ 2$\times$10$^{11}$M$_{\odot}$ & 0.5 $\leq z \leq$ 0.8 & 34 & 12 & 9  & 10 & 3 & 2.22 & 3.94 & 6.81 \\
 \hline
\multicolumn{10}{c}{Total MSFG sample}\\ 
 \hline
  All   & 0.8 $\leq z \leq$ 1.0 & 533 & 134 & 133 & 133 & 133 & 1.28 & 2.54 & 4.27 \\
  $\cal{M}$ $\leq$ 2$\times$10$^{11}$M$_{\odot}$ & 0.8 $\leq z \leq$ 1.0 & 477 & 120 & 119  & 119 & 119 & 1.22 & 2.46 & 4.28 \\
$\cal{M}$ $>$ 2$\times$10$^{11}$M$_{\odot}$ & 0.8 $\leq z \leq$ 1.0 & 56 & 14 & 14  & 14 & 14 & 2.00 & 3.11 & 4.27 \\
\hline
\multicolumn{10}{c}{Low-$\Sigma$ MSFG sample}\\ 
 \hline
    All         & 0.8 $\leq z \leq$ 1.0 & 398 & 96 & 105 & 95 & 102 & 1.28 & 2.54 & 4.27  \\  
  $\cal{M}$ $\leq$ 2$\times$10$^{11}$M$_{\odot}$ & 0.8 $\leq z \leq$ 1.0 & 360 & 84 & 96 & 85 & 95 & 1.22 & 2.46 & 4.28 \\
$\cal{M}$ $>$ 2$\times$10$^{11}$M$_{\odot}$ & 0.8 $\leq z \leq$ 1.0 & 38 & 10 & 10 & 10 & 8 & 2.00 & 3.11 & 4.27 \\
 \hline
\hline
   \end{tabular} \\
      \end{center}
   {\small   \textit{Column 1:} Sample. \textit{Column 2:} Redshift range. \textit{Column 3:} Total number of galaxies. \textit{Columns 4, 5, 6, \textup{and} 7:} Number of galaxies in the considered sample in the four $\delta$ bins defined according to the 25th$^{}$, 50th,$^{}$ and 75th$^{}$ percentiles of the (1+$\delta$) distribution. The values of these percentiles are listed in Cols. 8, 9, and 10 for each of the considered samples.}
     \label{sample}
 \end{table*}

Of the 902 MPGs in the final sample, 386 have $\Sigma \leqslant$\,1000\,M$_{\odot}$\,pc$^{-2}$ and 255 have $\Sigma >$\, 2000M$_{\odot}$\,pc$^{-2}$. The low- (high-) $\Sigma$ subsample can be split into four $\delta$ bins with at least $\sim$ 80(56) galaxies per bin (see Table\,\ref{sample}). 

\section{Galaxy size $\text{versus}$ environment at 0.5 < z < 0.8}\label{section3}

\begin{figure}
\includegraphics[width=9.5cm]{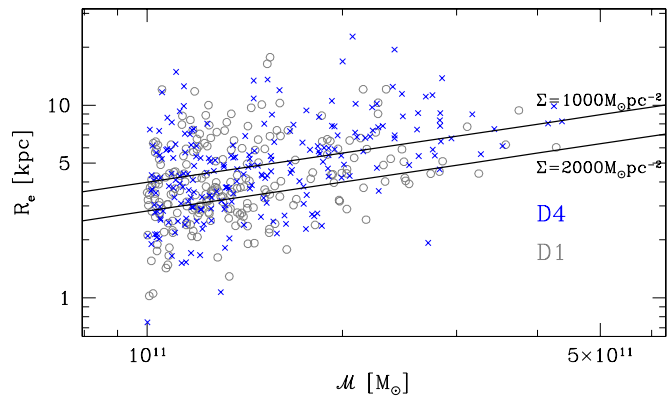}
        \caption{Size-mass relation of MPGs in the lowest (gray open circles) and highest (blue crosses) quartile of (1+$\delta$) distribution (D1 and D4, respectively). The solid lines show the constant mean stellar mass density $\Sigma,$ as annotated.}
    \label{sm}
\end{figure}

In Fig.\ref{sm} we plot the size-mass relation (SMR) of MPGs in the two extreme environments (i.e., MPGs with (1+$\delta$) $\leqslant$ 1.94 and (1+$\delta$) $\geqslant$ 6.48 (D1 and D4 bins, respectively).
The figure qualitatively shows that MPGs in the lowest and highest $\delta$ quartiles populate the same locus in the plane. This means that at fixed stellar mass, MPGs with a given $\Sigma$ can be found in any environment. In Appendix B we verify that this result is not dependent on our choice of the $\delta$ values used to identify low- and high-density regions. 

This does not tell us whether a given environment can instead favor the formation of a MPG with a given $\Sigma$, however. To address this point, we adopted the following approach. We compared 
the number of low- and high-$\Sigma$ MPGs in the four local density bins. By construction, each $\delta$ bin contains the same number of MPGs, therefore a constant number of the subpopulations of low- and high-$\Sigma$ MPGs as a function of $\delta$ implies no correlation between $\Sigma$ and $\delta$.

Before we carried this comparison out, we had to apply a correction for the survey incompleteness. VIPERS has three sources of incompleteness: the TSR, the success sampling rate (SSR), and the color sampling rate (CSR). As stated in Sect. 2, the TSR is the fraction of galaxies that are effectively observed with respect to the photometric parent sample. The SSR is the fraction of spectroscopically observed galaxies with a redshift measurement. The CSR accounts for the incompleteness due to the color selection of the survey. These statistical weights depend on the magnitude of the galaxy, on its redshift, color, and angular position. They have been derived for each galaxy in the full VIPERS sample \citep[for a detailed description of their derivation see][]{garilli14, scodeggio16}.  To correct the number of MPGs in each $\delta$ bin for all of these incompletenesses, we weighted each galaxy $i$ by the quantity \textit{w$_i$} = 1/(\textit{TSR$_i$}$\times$\textit{SSR$_i$}$\times$\textit{CSR$_i$}). 
We do not expect these sources of incompleteness to affect our results because the galaxy size was not considered in the slit assignment in VIPERS. On the other hand, the SSR could be higher for dense MPGs with respect to less dense MPGs because their light profile is more strongly peaked. However, this higher SSR does not depend on the environment because it is only related to the intrinsic properties of the galaxies. The same holds for the CSR. This ensures that the populations of MPGs in dense and less dense environment are not biased in $\Sigma$.

\begin{figure}
        \includegraphics[width=9.4cm]{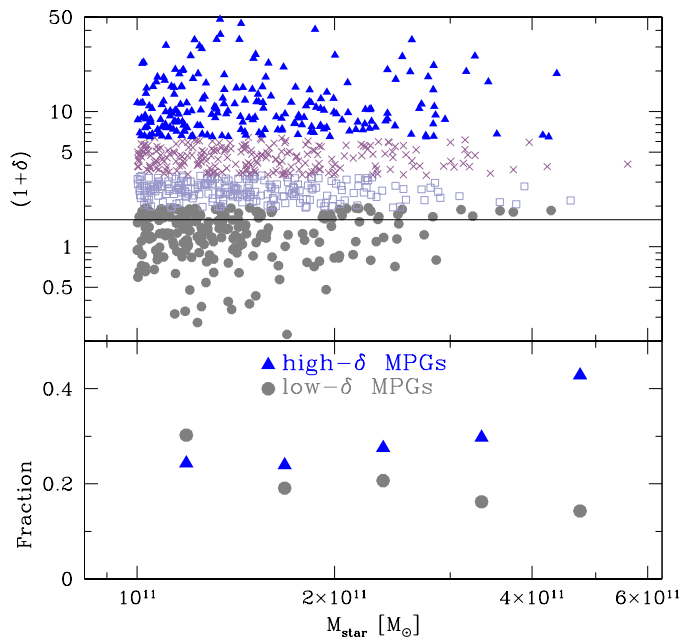} 
    \caption{\textit{Upper panel}: Local density $\delta$ as a function of galaxy stellar mass for our MPG sample. Filled gray circles, open violet squares, pink crosses, and filled blue triangles indicate MPGs in the D1, D2, D3, and D4 quartiles. The horizontal solid line indicates the $\delta$ value above which MPGs cover the full range of stellar mass ($1+ \delta = 1.5$) (see~Sect.~\ref{section3}). \textit{Bottom panel}:  fraction of MPGs in the highest and lowest $\delta$ quartiles as a function of galaxy stellar mass (blue triangles and gray full circles, respectively). }
    \label{trendmass}
        \end{figure}

Another factor that can bias our results is the well-known
correlation between $\cal{M}$ and $\delta,$ whereby more massive
galaxies populate denser environments.  We examine the stellar
mass distribution of our sample of MPGs as a function of $\delta$ in the
upper panel Fig.\,\ref{trendmass}.  
The figure shows that the most massive galaxies are underrepresented in the low-density bin.  We quantify this difference in the bottom panel of Fig.\,\ref{trendmass}, where the fraction of MPGs in the lowest and highest $\delta$ quartiles is reported as a function of the stellar mass.  In agreement with previous results, we find that the fraction of MPGs in the densest regions increases (by a factor $\sim$2) for stellar masses increasing from 10$^{11}$\,M$_{\odot}$ to 5$\times$10$^{11}$\,M$_{\odot}$, and concurrently, the fraction of MPGs in the less dense regions decreases with ${\cal M}$.  

Figure~\ref{sm} shows that at higher stellar masses, low-$\Sigma$ MPGs prevail over the full MPG population, resulting in an increasing fraction of low-$\Sigma$ MPGs with $\cal{M}$.  This trend,
coupled with the $\cal{M}$-$\delta$ correlation shown in Fig.\,\ref{trendmass}, could bias our results toward a false trend of larger galaxies in denser environment. 

To remove the possible biases that are due to the different stellar mass distributions of MPGs in low- and high-density regions, we extracted four subsamples of MPGs with the same distribution in stellar mass  from
the four $\delta$ bins (see Appendix C).
This reduced the total number of MPGs to 165 galaxies in each local density quartile.

\begin{figure}
\includegraphics[width=9.1cm]{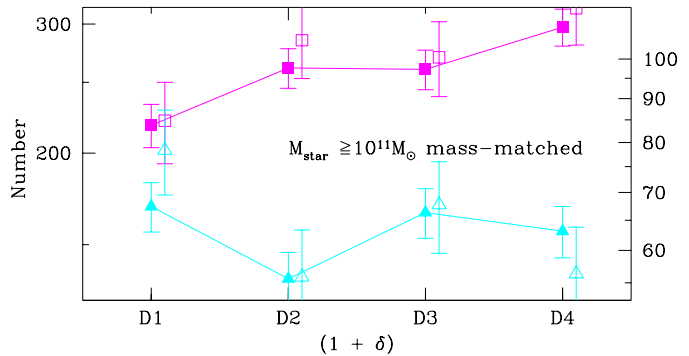}
        \caption{Average number of low-$\Sigma$ (filled magenta squares) and high-$\Sigma$ (filled cyan triangles) MPGs as a function of local density $\delta$ derived from 100 mass-matched samples. The counts are corrected for the selection function of the VIPERS survey.  Error bars include the Poisson fluctuations and the error on the mean. The four $\delta$ bins correspond to the four quartiles of the MPG (1+$\delta$) distribution. Open symbols are the mean number of low- and high-$\Sigma$ MPGs that are effectively observed (i.e., not corrected for the selection function) over 100 iterations.  The ordinate  scale on the right-hand side refers to the uncorrected counts. Light shaded magenta and cyan areas indicate the $\pm$1$\sigma$ region around the mean value of low- and high-$\Sigma$ MPGs that are expected in each bin when no trend with $\delta$ is assumed.}
    \label{nummm}
\end{figure}

The filled magenta squares and filled cyan triangles in Fig.\,\ref{nummm}  indicate the mean number of low- and high-$\Sigma$ MPGs in each $\delta$ bin derived from the 100 mass-matched samples. Error bars take  the Poisson fluctuations and the error on the mean into account. Open symbols  indicate the same values, but with no correction factor for incompleteness.
For both low- and high-$\Sigma$ MPGs, the trend with $\delta$ that is obtained considering the corrected estimates is consistent with the trend obtained with uncorrected estimates. This indicates that the weights we applied do not alter the trends, and it confirms that none of the incompleteness factors depends on $\delta$ or $\Sigma.$  For completeness, we report in Appendix D the same as in Fig.  \ref{nummm}, but for the original samples, that is, not mass matched. In the following we refer only to the results obtained with corrected values (filled points in Fig.\,\ref{nummm}).

Figure \ref{nummm} shows similarly to Fig.~\ref{sm} that low- and
high-$\Sigma$ MPGs are ubiquitous in all environments. In addition, it shows that the number of high-$\Sigma$ MPGs is almost constant over the four $\delta$ bins ($\chi^{2}$ = 3.3, the probability $P$ that $\chi^{2} \geqslant$ 3.3 is 0.35). In particular, the number of high-$\Sigma$ MPGs in each $\delta$ bin is consistent within 1$\sigma$ with the mean number derived when no trend with $\delta$ is assumed. 

The number of low-$\Sigma$ MPGs slightly and steadily increases from the lowest to the highest $\delta$ quartiles with a $\chi^{2}$ = 8.6 and a probability P($\chi^{2} \geqslant$ 8.6) = 0.035. In particular, the number of low-$\Sigma$ MPGs in the D1 region is more than 1$\sigma$ lower than the number of low-$\Sigma$ MPGs in the D4 region.
The difference in the $\delta$ distributions of low- and high-$\Sigma$ MPGs was compared through a Kolmogorov-Smirnov (KS) test. From each mass-matched sample we selected the low- and high-$\Sigma$ MPGs and derived the KS probability $p$ that their $(1+\delta)$ distributions were extracted from the same parent sample. 
The median value of $p$ over the 100 simulations is 0.11, suggesting that we cannot exclude that the parent population of the two distributions are the same single population.
Although the trends we found are marginally significant, they might imply that galaxy size and environment are also related at redshifts 0.5--0.8, and thus it is important to investigate it further.  

\subsection{Dependence of the $\delta$-$\Sigma$ relation on stellar mass}

Several relations involving the stellar mass (e.g., the SMR) show a sharp change in their trends at $\cal{M}$ $\sim$ 2$\times$10$^{11}$M$_{\odot}$ \citep[e.g.,][]{cappellarixx, bernardi11a, bernardi11b,saracco17}. This feature is interpreted as an indication of an increasing role of satellite accretion in the mass-assembly history of galaxies with stellar mass above such a threshold \citep{bernardi11a,bernardi11b}. According to this picture, we might expect that MPGs with $\cal{M}$ $>$ 2$\times$10$^{11}$M$_{\odot}$ show a stronger positive correlation between their size and the environment than the less massive ones. To test this prediction, we investigated whether the higher number of low-$\Sigma$ MPGs we observe in denser environment is mainly due to galaxies with $\cal{M}$ $>$ 2$\times$10$^{11}$M$_{\odot}$.

\begin{figure}
\includegraphics[width=9.1cm]{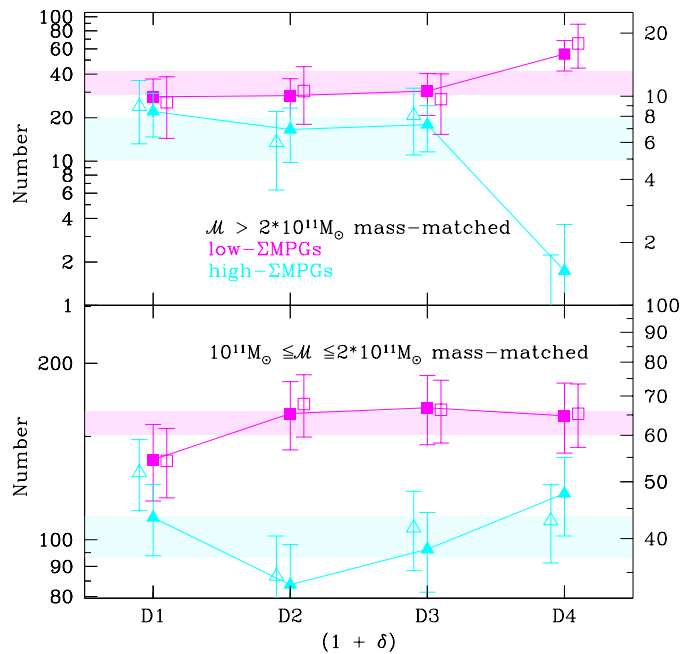}
        \caption{Same as Fig.~\ref{nummm}, but for  MPGs with 10$^{11}$M$_{\odot}$ $\leqslant \cal{M} \leqslant$ 2$\times$10$^{11}$M$_{\odot}$ (lower panel) and $\cal{M} >$2$\times$10$^{11}$M$_{\odot}$ (upper panel).}
    \label{nummassmm}
\end{figure}

In Fig.\,\ref{nummassmm} we report the same analysis as in Fig.\,\ref{nummm}, but now for MPGs with 10$^{11}$M$_{\odot}$ $\leqslant \cal{M} \leqslant$ 2$\times$10$^{11}$M$_{\odot}$ (lower panel), and with $\cal{M} >$2$\times$10$^{11}$M$_{\odot}$ (upper panel).
The figure shows that, no significant difference in the number of high- and low-$\Sigma$ MPGs with $\delta$ is detectable at lower stellar masses (bottom panel; all the values are consistent at $<1\sigma$ level). 
At $\cal{M} >$2$\times$10$^{11}$M$_{\odot}$, the numbers of both high- and low-$\Sigma$ MPGs are all consistent  in the first three $\delta$  bins, but in the last bin, the number of low-$\Sigma$ MPGs suddenly increases by a factor $\sim$2, while the number of high-$\Sigma$ MPGs drastically drops by a factor $\sim$ 10. The D4 region appears to be lacking high-$\Sigma$ MPGs with $\cal{M} >$2$\times$10$^{11}$M$_{\odot}$, and to have an overabundance of low-$\Sigma$ MPGs with the same stellar mass.

\begin{figure}
\includegraphics[width=9.5cm]{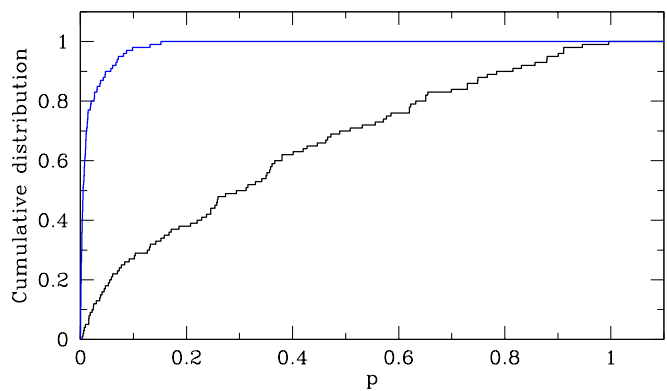}
        \caption{Cumulative distribution of the KS-probability $p$ that the $(1+\delta)$ distributions of low- and high-$\Sigma$ MPGs are extracted from the same parent population that is obtained from all of the 100 mass-matched samples. The black line refers to MPGs with 10$^{11}$M$_{\odot}$ $\leqslant \cal{M} \leqslant$ 2$\times$10$^{11}$M$_{\odot}$, and the blue line to MPGs with $\cal{M} >$ 2$\times$10$^{11}$M$_{\odot}$. }
    \label{massmatchp}
\end{figure}

The different $\Sigma$-$\delta$ trends in the two stellar mass bins is also evident in figure \ref{massmatchp}, where we report for the two stellar mass bins the cumulative distribution of the  KS-test probability $p$ that the $(1+\delta)$ distributions of low- and high-$\Sigma$ MPGs from the 100 mass-matched samples are extracted from the same parent population. The figure shows that the median value of $p$ for the high-mass sample is 5$\times$10$^{-3}$, and for the low-mass sample is $\sim$0.3. Figures \ref{nummassmm} and\,\ref{massmatchp} indicate that the dependence on the environment of low- and high-$\Sigma$ MPGs with $\cal{M} >$ 2$\times$10$^{11}$M$_{\odot}$ is significantly different, with a higher probability to find large (low-$\Sigma$) galaxies in denser environment. Conversely, we do not detect any significant difference in the probability of finding a low- or a high-$\Sigma$ MPG with 10$^{11}$M$_{\odot}$ $\leqslant \cal{M} \leqslant$ 2$\times$10$^{11}$M$_{\odot}$ in a specific environment.

Very few works have investigated the relation between the size/$\Sigma$ and the environment in the same redshift and mass ranges as in this work. \citet{kelkar15} find no significant differences in the ${\it R}_e$ of red galaxies with $\cal{M} >$ 10$^{11}$M$_{\odot}$ in clusters and fields at $0.4 < z < 0.8$. This result is partially at odds with our findings because we observe a correlation at the highest mass-scales, but this discrepancy is entirely accounted for by the different observables used in the two analyses. Using their samples, we selected red galaxies on the basis of their $\Sigma$ instead of their ${\it R}_e$ and found that the number of low-$\Sigma$ galaxies with $\cal{M} >$ 2$\times$10$^{11}$M$_{\odot}$ in clusters is more than twice the number of low-$\Sigma$ galaxies in the field (9 vs. 4, respectively). This qualitatively agrees with our findings (the number of low-$\Sigma$ MPGs in D4 regions almost doubles the number of low-$\Sigma$ MPGs in D1 regions in this mass range, as shown in the top panel of Fig.\,\ref{nummassmm}). Similarly, \citet{lani13} measured the local galaxy density in fixed physical apertures of radius 400 kpc and found no dependence between size and environment of quiescent (UVJ selected) galaxies at $0.5 < z < 1.0$ with $\cal{M} \sim$ 10$^{11}$M$_{\odot}$. At $\cal{M} >$2$\times$10$^{11}$M$_{\odot}$ they found no dependence between the ${\it R}_e$ and the environment except for galaxies in the densest regions that are larger than those in other environments. These trends are consistent with the results shown in Fig.\,\ref{nummassmm}. Differently from our findings, \citet{huertascompany13} found no dependence of galaxy size on environment, regardless of stellar mass. In their work, they investigated the ${\it R}_e$ - $\delta$ relation for passive $\text{and}$ elliptical galaxies, differently from our analysis, which is based on color-selected passive galaxies (as in \citet{lani13} and \citet{kelkar15}). It is well known that samples of passive galaxies selected through colors contain $20-30\%$ disk galaxies \citep[e.g.,][]{huertascompany13,tamburri14,moresco13}. The different selection method might explain the difference between our MPG sample and their passive galaxy sample, but we are not in the position to quantify the effect of this difference on the results.

\subsection{Relation between $\delta$ and $\Sigma$ for MPGs: summary $\text{and}$ discussion}

The results we obtained for the mass-matched samples and presented in 
Figure \ref{nummassmm} indicate that
\begin{itemize}
\item at 10$^{11}$M$_{\odot}$ $\leqslant \cal{M} \leqslant$ 2$\times$10$^{11}$M$_{\odot}$ we do not observe a significant correlation between the $\Sigma$ of a MPG and the environment in which it lives;
\item at $\cal{M} >$2$\times$10$^{11}$M$_{\odot}$ we find a flat trend between the number of low- and high-$\Sigma$ MPGs and $\delta$ in the first three $\delta$ bins. In the highest density bin, the number of low-$\Sigma$ MPGs instead suddenly increases by $\sim$ a factor 2 with respect to the bins at lower density, while the number of high-$\Sigma$ MPGs decreases by a factor $\sim$ 10. 
\end{itemize}

These results indicate that there is no environment that favors (or disfavors) the formation of low- and high-$\Sigma$ MPGs with 10$^{11}$M$_{\odot}$ $\leqslant \cal{M} \leqslant$ 2$\times$10$^{11}$M$_{\odot}$, that is, the mechanisms favored in denser environments, such as satellite accretion or cannibalism, do not appear to be responsible for the low mean surface stellar mass density of large MPGs. 

The picture is more complex at higher stellar mass. We observe a decline in the number of high-$\Sigma$ MPGs in the densest environment and concurrently an increase in the number of the low-$\Sigma$ MPGs. This suggests two main scenarios: $i$) the densest environments favor the formation of low-$\Sigma$ MPGs with $\cal{M} >$2$\times$10$^{11}$M$_{\odot}$ and concurrently disfavor the formation of high-$\Sigma$ MPGs with the same stellar mass; $ii$) low- and high-$\Sigma$ MPGs are formed in all environments with the same probability, that is, the formation (at $t=t_0$) of dense or not dense passive galaxies does not depend on $\delta$ as for less massive counterparts; but at a later time ($t_1 > t_0$), high-$\Sigma$ MPGs disappear in the most dense regions and low-$\Sigma$ MPGs appear. This means that their $\text{evolution}$ does depend on the environment. 
The hierarchical accretion of satellite galaxies around a compact galaxy could be the driver of a migration of MPGs from the high-$\Sigma$ sample toward the low-$\Sigma$ sample, and consequently be the reason of the abundance (deficit) of low- (high-) $\Sigma$ MPGs  with respect to the other $\delta$ bins. 
Predictions of standard hierarchical models \citep[e.g.,][]{shankar12, huertascompany13} indicate that at these mass scales, galaxies in groups should be $\sim$ 1.5 times larger than galaxies in less dense fields. We have estimated the mean ${\it R}_e$ of MPGs with $\cal{M} >$2$\times$10$^{11}$M$_{\odot}$ in the four density bins and found that it is $\sim$\,1.4 times larger in the D4 bin than in the other less dense environments ($\langle {\it R}_e \rangle$ = 6.24$\pm$0.04, 6.18$\pm$0.03, 6.21$\pm$0.03, and 8.73$\pm$0.08\,kpc for the D1, D2, D3, and D4 bins, respectively).  The concordance between hierarchical model predictions and our observations indicates that  accretion of satellites could be a viable mechanisms for the formation and evolution of those largest (e.g., less dense) MPGs with $\cal{M} >$2$\times$10$^{11}$M$_{\odot}$ in excess in the densest regions. This supports the second scenario described above.

Another way to distinguish between the two scenarios would be to investigate the $\Sigma-\delta$ relation for MPGs with $\cal{M} >$2$\times$10$^{11}$M$_{\odot}$ at higher $z$. In the first scenario, we should observe no evolution of the $\Sigma$-$\delta$ relation with time. Conversely, in the second picture, we should observe a flat $\Sigma$-$\delta$ trend at $t=t_0$ because the correlation between $\Sigma$ and $\delta$ observed in the top panel of Fig.\,\ref{nummassmm} should appear later, after satellite accretion has become relevant. 

Unfortunately, MPGs with $\cal{M} >$ 2$\times$10$^{11}$M$_{\odot}$ at $z > 0.8$ are very rare in VIPERS. In spite of the wide coverage of survey, we find only 46 MPGs in the redshift range $0.8 < z < 0.9$. This prevents an analysis at higher $z$ (we highlight that these 46 MPGs have to be divided into four $\delta$ bins and into low- and high-$\Sigma$ subsamples). Nonetheless, the VIPERS sample offers a unique statistics at $z \geqslant 0.8$ for MSFGs. In Sect. 4 we use the sample of MSFGs at $ 0.8 \leqslant z \leqslant 1.0$ as a benchmark for testing our conclusions.

\section{Number of MSFGs as a function of $\delta$}
\label{ndsec}
The analysis shown in Fig.\,\ref{nummassmm} 
questions the relevance of satellite accretion in the mass assembly of low-$\Sigma$ MPGs with 10$^{11}$M$_{\odot}$ $\leqslant \cal{M} \leqslant$ 2$\times$10$^{11}$M$_{\odot}$. Using a different probe (e.g., the number density evolution of MPGs and MSFGs), we reached the same conclusion in Paper I, where we 
suggested that  MSFGs at $z \geqslant 0.8$ are the direct progenitors of low-$\Sigma$ MPGs. 
If the interpretation we draw based on these two works is correct, we should find that  the low-$\Sigma$ MSFGs at $z \geqslant 0.8$ follow the same trend with $\delta$ as  low-$\Sigma$ MPGs at $z <$ 0.8 because the former are the progenitors of the latter. To test this hypothesis, we repeated the same analysis as in Sect. 3 for the sample of MSFGs at 0.8 $\leqslant z \leqslant$ 1.0. In this analysis we refer to the density field $\delta$ that is estimated using VIPERS galaxies with M$_{B} \leqslant$ (-20.9 - $z$)  as tracers (see Sec. 2.1). Although this density field is computed on larger scales than the field used for MPGs,  it allows us to push the analysis up to $z = 1.0$. In Appendix E we show that this choice does not affect our results and conclusions.

In Table\,\ref{sample} we report the total number of MSFGs at 0.8 $\leqslant z \leqslant$ 1.0 with the values of the 25th, 50th, and 75th percentiles of their (1+$\delta$) distribution. As for MPGs, we also report the number of galaxies (and of percentiles) for the two subsamples with 10$^{11}$M$_{\odot}$ $\leqslant \cal{M} \leqslant$ 2$\times$10$^{11}$M$_{\odot}$ and $\cal{M} >$2$\times$10$^{11}$M$_{\odot}$.

\begin{figure}
\includegraphics[width=9.0cm]{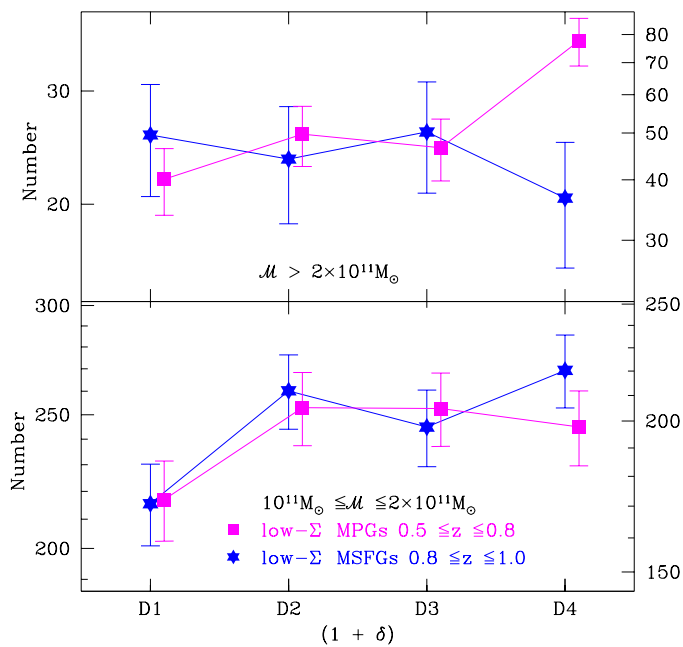}
        \caption{\textit{Lower panel}: Number of low-$\Sigma$ MSFGs at 0.8 $\leqslant z \leqslant$ 1.0 (blue stars) with 10$^{11}$M$_{\odot}$ $\leqslant \cal{M} \leqslant$ 2$\times$10$^{11}$M$_{\odot}$ as a function of local density $\delta$.  For comparison, we also report the number of  low-$\Sigma$ MPGs at 0.5 $\leqslant z \leqslant$ 0.8 (magenta squares) with 10$^{11}$M$_{\odot}$ $\leqslant \cal{M} \leqslant$ 2$\times$10$^{11}$M$_{\odot}$. Their values are reported on the right y-axis. The four $\delta$ bins correspond to the four quartiles of the MPG (1+$\delta$) distribution. \textit{Upper panel}: Same as in the lower panel, but for MPGs with $\cal{M} >$ 2$\times$10$^{11}$M$_{\odot}$}
    \label{numdensf}
\end{figure}

In the bottom panel of Fig.\,\ref{numdensf} we report the number of low-$\Sigma$ MSFGs at 0.8 $\leqslant z \leqslant$ 1.0  with 10$^{11}$M$_{\odot}$ $\leqslant \cal{M} \leqslant$ 2$\times$10$^{11}$M$_{\odot}$ as a function of $\delta$. For comparison, we also report the number of low-$\Sigma$ MPGs at 0.5 $\leqslant z \leqslant$ 0.8, and in the same mass range. In this case, we did not use the mass-matched values, but the values that were just corrected for incompleteness because we are interested in the relative change with $\delta$ of the two populations.
The figure shows that the two trends are consistent at the 1$\sigma$ level. This result, combined with the results in Paper I and Sect. 3, strengthens the postulated evolutionary connection between MSFGs and MPGs.
In Paper I we demonstrated that MSFGs at $z \sim 0.8$ are in the correct number to be the progenitors of low-$\Sigma$ MPGs at $z < 0.8$, and here we show that they are also in the correct environment. The analyses of the number density and of the environment concur in indicating that low-$\Sigma$ MPGs with 10$^{11}$M$_{\odot}$ $\leqslant \cal{M} \leqslant$ 2$\times$10$^{11}$M$_{\odot}$  at $z < 0.8$ are the descendants of MSFGs at higher $z$ and not the result of  satellite accretion.

In Sect. 3.2 we have shown that the accretion of satellites could be a viable mechanisms for the formation of the exceedingly low-$\Sigma$ MPGs with $\cal{M} >$ 2$\times$10$^{11}$M$_{\odot}$ in the D4 bin.
In this picture, the $\Sigma$-$\delta$ trend of low-$\Sigma$ MPGs at $z < 0.8$ should not be consistent with that of MSFGs at $0.8 < z < 1.0$, as we found for lower mass-scales. The two trends are expected to deviate in the highest density bin because the majority of MSFGs become passive at $z < 0.8,$ but to these "shutdown" MPGs we should add in the highest density bin the large MPGs that are formed by the accretion of satellites.

In the top panel of Fig.\,\ref{numdensf} we report the $\Sigma$ - $\delta$ trend for MSFGs with $\cal{M} >$ 2$\times$10$^{11}$M$_{\odot}$. It shows that in the highest stellar mass bin, the $\Sigma$ - $\delta$ trend for MSFGs at $z > 0.8$ is flat (all points are consistent within 1$\sigma$).  We observe instead a flat $\Sigma$ - $\delta$ trend for MPGs at $z < 0.8$ in the first three $\delta$ bins and an increase in the number of low-$\Sigma$ MPGs in the highest density bin (see the magenta points in the upper panel of Fig.\,\ref{numdensf}). 
This evidence supports the second scenario we proposed in Sect. 3.2. The flat trend of MSFGs over the four $\delta$ bins indicates that at $t=t_0$, the probability that a MSFG with $\cal{M} >$ 2$\times$10$^{11}$M$_{\odot}$ (and consequently of its descendant MPG) has a low-$\Sigma$ is independent of the environment. At $t_1 > t_0$, environment-related mechanisms, such as accretion of satellites, are responsible for the surplus of MPGs that is observed in the D4 bin.

 In summary, our analysis shows that the larger size of {\it \textup{most of the}} low-$\Sigma$ MPGs is not a consequence of a hierarchical assembly of the stellar matter. Although this is true for the majority of the galaxies in our sample, we cannot rule out satellite accretion as one of the main mechanisms responsible for increasing the size, thus decreasing $\Sigma$, in a small fraction of high-$\Sigma$ MPGs with $\cal{M} >$2$\times$10$^{11}$M$_{\odot}$ that reside in the densest environments (i.e., $\sim$ 10 high-$\Sigma$ MPGs over $\sim$902, where 10 is the difference between the mean value of high-$\Sigma$ MPGs in the D1, D2, and D3 bins and the value in the D4 bin, see the upper panel of Fig. \ref{nummassmm}, and 902 is the total number of MPGs). 

\section{Summary}

Using the VIPERS spectroscopic survey, we investigated the relation between the typical size of MPGs and their environment. In a hierarchical scenario, structures are expected to grow in mass and size through accretion of smaller units. 
The accretion of satellites is enhanced in groups (and for central galaxies), thus we would expect to find larger galaxies in groups than in less dense environments. The study of galaxy sizes as a function of the environment therefore is a powerful diagnostic for galaxy accretion models.

From the VIPERS database we extracted all the galaxies at 0.5 $\leqslant z \leqslant$ 0.8 with $\cal{M} \geqslant$ 10$^{11}$M$_{\odot}$. From these systems we selected the passive galaxies according to their NUV$r$K colors. The sample is 100$\%$ complete in stellar mass at $z = 0.8.$  
From the massive and passive galaxies, we selected those with reliable estimates of the effective radius $\it{R_e}$ and of the galaxy density contrast $\delta$. This led to a total final sample of 902 MPGs in the redshift range 0.5 $\leqslant z \leqslant$ 0.8.

From this sample, we selected a low-$\Sigma$ subsample composed of MPGs with a mean surface stellar mass density $\Sigma \leqslant$ 1000\,Mpc\,pc$^{-2}$, and a high-$\Sigma$ subsample that contains the MPGs with $\Sigma >$ 2000\,Mpc\,pc$^{-2}$.

The overall picture we obtained from our analysis of size $\text{versus}$ environment in VIPERS galaxies at $0.5 \leqslant z \leqslant 0.8$ is described below.
\begin{itemize}
\item The $\Sigma$-$\delta$ relation depends on stellar mass. In particular, MPGs with 10$^{11}$M$_{\odot}$ $\leqslant \cal{M} \leqslant$ 2$\times$10$^{11}$M$_{\odot}$ do not show any significant trend between the $\Sigma$ (or equivalently $\it{R_e}$) and the environment (the KS probability $p$ that the density contrast distributions of low- and high-$\Sigma$ MPGs are extracted by the same parent sample is 0.3, see Fig.\,\ref{massmatchp}), while a significant correlation is detected for MPGs with $\cal{M} >$ 2$\times$10$^{11}$M$_{\odot}$ ($p = 5\times10^{-3}$), with larger galaxies being more common in denser environments. 
\item The trend between $\Sigma$ and $\delta$ observed for MPGs with $\cal{M} >$ 2$\times$10$^{11}$M$_{\odot}$ is due to an overabundance of low-$\Sigma$ MPGs and by a dearth of high-$\Sigma$ MPGs in the densest regions (see Fig.\,\ref{nummassmm}).
\end{itemize}

The absence of a correlation between $\Sigma$ and $\delta$ for MPGs with 10$^{11}$M$_{\odot}$ $\leqslant \cal{M} \leqslant$ 2$\times$10$^{11}$M$_{\odot}$ indicates that the accretion of satellite galaxies is not the main driver of the build-up of low-$\Sigma$ MPGs in this stellar mass range. This result is in line with our previous results. In Paper I (see Fig. 9), the analysis of the number densities of MPGs and MSFGs suggested that the most plausible progenitors of the emerging low-$\Sigma$ MPGs at $z < 0.8$ are the MSFGs at $z > 0.8$. We further investigated this connection and in Fig.\,\ref{numdensf} compared the $\Sigma - \delta$ trends for MSFGs at $z > 0.8$ and low-$\Sigma$ MPGs $z < 0.8$. The two trends are consistent at the 1$\sigma$ level, as it should be if the two populations are evolutionary connected. In Paper I we demonstrated that the MSFGs are in the correct number to be the progenitors of low-$\Sigma$ MPGS, and here we show that they are also in the correct environment.

For the highest stellar mass bin, our results on the $\Sigma$-$\delta$ correlation of both MPGs and MSFGs are consistent with a scenario in which, overall, environment does not affect the $\text{formation}$ of low- or high-$\Sigma$ MPGs. Nonetheless, it seems to play a relevant role in the $\text{evolution}$ of a small fraction ($< 1\%$) of low-$\Sigma$ MPGs in the densest environment. We interpret the increased number of low-$\Sigma$ MPGs in the densest regions as a consequence of a migration of high-$\Sigma$ MPGs in low-$\Sigma$ MPGs that is due to accretion of satellite galaxies. These results are also confirmed by comparison with predictions of hierarchical models.

\begin{acknowledgements}
We acknowledge the crucial contribution of the ESO staff for the management of service observations. In particular, we are deeply grateful to M. Hilker for his constant help and support of this program. Italian participation to VIPERS has been funded by INAF through PRIN 2008, 2010, and 2014 programs. LG, AJH, and BRG acknowledge support from the European Research Council through grant n.~291521. OLF acknowledges support from the European Research Council through grant n.~268107. TM and SA  acknowledge financial support from the ANR Spin(e) through the French grant  ANR-13-BS05-0005. AP and JK have been supported by the National Science Centre (grants UMO-2018/30/M/ST9/00757 and
UMO-2018/30/E/ST9/00082). KM acknowledges support from the National Science Centre grant UMO-2018/30/E/ST9/00082. WJP is also grateful for support from the UK Science and Technology Facilities Council through the grant ST/I001204/1. EB, FM and LM acknowledge the support from grants ASI-INAF I/023/12/0 and PRIN MIUR 2010-2011. SDLT acknowledges the support of the OCEVU Labex (ANR-11-LABX-0060) and the A*MIDEX project (ANR-11-IDEX-0001-02) funded by the "Investissements d'Avenir" French government program managed by the ANR and the Programme National Galaxies et Cosmologie (PNCG). Research conducted within the scope of the HECOLS International Associated Laboratory is supported in part by the Polish NCN grant DEC-2013/08/M/ST9/00664.
\end{acknowledgements}

\nocite{}
\bibliographystyle{aa}
\bibliography{paper_II}

\begin{appendix}
 \appendix

\section{ $\delta$ distribution as a function of $z$}
\label{appa}
\begin{figure}
\includegraphics[width=9.5cm]{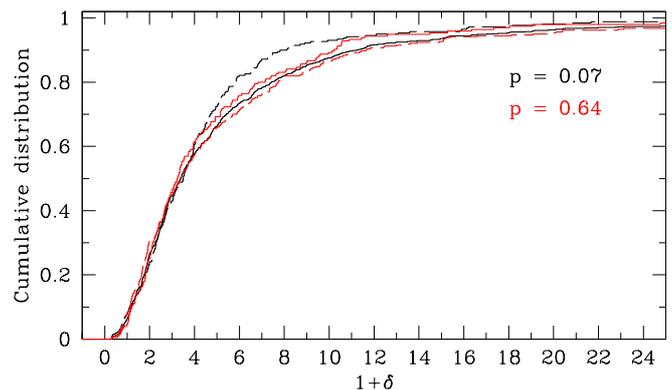}
        \caption{Cumulative distribution of (1+$\delta$) for MPGs with reliable $\it{R_e}$ and $\delta$ at 0.5 $\leqslant z \leqslant$ 0.8 (black solid line) and at 0.8 $< z \leqslant$ 0.9 (dashed black line). The probability $p$ that the two distributions are extracted from the same parent sample is rejected at $\sim$3$\sigma$. Red solid and dashed lines indicate the cumulative distribution for MPGs at 0.5 $\leqslant z \leqslant$ 0.6 and 0.7 $\leqslant z \leqslant$ 0.8, respectively. The probability that they are extracted from the same parent sample is 0.64. }
    \label{deltacum}
\end{figure}

In this section we investigate the dependence of the (1+$\delta$) distribution of MPGs on $z$. In Fig.\,\ref{deltacum} we plot the cumulative distributions of the local density for MPGs with reliable $\it{R_e}$ and $\delta$ in the two different redshift bins 0.5 $\leqslant z \leqslant$ 0.8 and 0.8 $< z \leqslant$ 0.9. The KS test indicates a probability $p$ = 0.07 that the two distributions are extracted from the same parent sample. Conversely, we do not find a strong dependence between $z$ and the  $\delta$ distributions at $z \leqslant 0.8$. As an example, the red lines in Fig.\,\ref{deltacum} indicate the  (1+$\delta$) distribution for MPGs in the redshift bins 0.5 $\leqslant z \leqslant$ 0.6 and 0.7 $\leqslant z \leqslant$ 0.8. The probability that the two distributions are extracted from the same parent sample is $p$ = 0.64.

\section{Effect of the environment on the stellar mass distribution}
\label{appb}

\begin{figure}
\includegraphics[width=9.5cm]{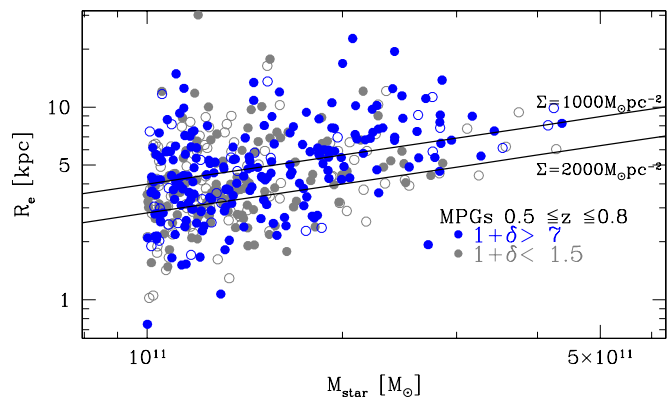}
        \caption{Size-mass relation of MPGs in the final sample with (1+$\delta$) $<$ 1.5 (gray filled circles) and with (1+$\delta$) $>$ 7 (blue filled circles). Open blue and gray circles are the MPGs that were excluded from Fig.\,\ref{sm}, i.e., MPGs with 1.5 $\leqslant (1+\delta) \leqslant$ 1.94 and 6.48 $\leqslant (1+\delta) \leqslant$ 7, respectively. }
    \label{smdiff}
\end{figure}

In this section we investigate the size-mass relation in two extreme environments. In particular,
we report the size-mass relation for MPGs with (1+$\delta$) $\leqslant$ 1.5 and  $(1 + \delta)$ $\geqslant$ 7 (see Fig.\,\ref{trendmass} and the text for the choice of these new cuts). 
We find that MPGs in regions with $(1 + \delta)$ $\geqslant$ 7 populate similar regions in the $\it{R_e}$ versus $\cal{M}$ plane of MPGs with $(1 + \delta)$ $\geqslant$ 6.48 (see the filled and open blue points in Fig.\,\ref{smdiff}). Conversely, the population of MPGs with (1+$\delta$) $\leqslant$ 1.5 does not include the most massive galaxies with respect to the population of MPGs with (1+$\delta$) $\leqslant$ 1.94. 
Despite the lack of massive galaxies in the lowest density regions, Fig.\,\ref{smdiff} shows that in the stellar mass range covered both by MPGs with $(1 + \delta)$ $<$ 1.5 and by MPGs with $(1 + \delta)$ $>$ 7 (i.e., $\cal{M} <$3$\times$10$^{11}$\,M$_{\odot}$) the range of $\it{R_e}$ that is covered is not significantly different. This indicates that environment plays an important role in setting the stellar mass (we take this aspect into account in Sect. 3.2), as is known, but it confirms that our results shown in Fig.\,\ref{sm} are robust against the choice of the $\delta$ cuts. 

\section{Construction of the mass-matched samples}

\begin{figure}
\includegraphics[width=9.5cm]{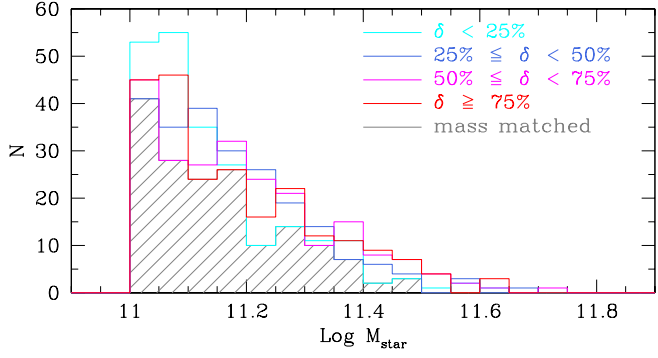}
        \caption{Stellar mass distribution of MPGs at 0.5 $\leqslant z \leqslant$ 0.8 in the four $\delta$ bins (cyan, blue, magenta, and red histograms for the D1, D2, D3, and D4 $\delta$ bin, respectively). The dashed gray histogram indicates the stellar mass distribution of the mass-matched sample.}
    \label{mmdelta}
\end{figure}

To construct the mass-matched samples we used in the analysis of Sect. 3, we started from the samples of MPGs in the four $\delta$ bins. We subdivided the four samples into bins of 0.05dex in stellar mass. For any stellar mass bin, we identified the sample with the lowest number of objects, and then randomly extracted from the others an equal number of galaxies. The stellar mass distributions of MPGs in the four $\delta$ bins and the stellar mass distribution of a mass-matched sample are shown in Fig.\,\ref{mmdelta}. 

\section{Number of MPGs as a function of $\Sigma$ and $\delta$ for the not mass-matched samples}

\begin{figure}
\includegraphics[width=9.1cm]{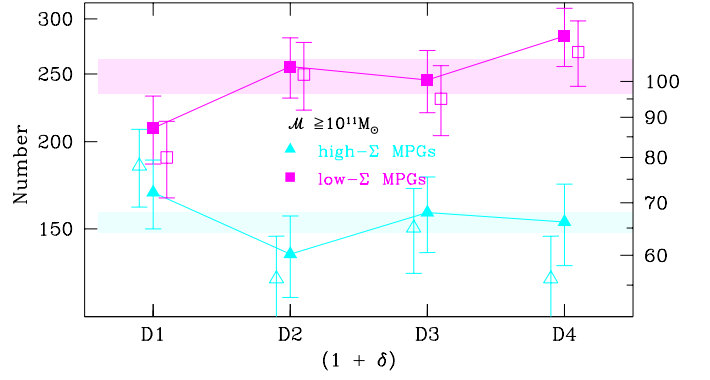}
        \caption{Same as Fig. 5, but now the filled magenta and cyan symbols show the number of low-$\Sigma$ and high-$\Sigma$ MPGs corrected for the selection function of the VIPERS survey. Error bars take the Poisson fluctuations into account. Open symbols show the number of low- and high-$\Sigma$ MPGs that are effectively observed (i.e., not corrected for the selection function).
        }
    \label{numnew}
\end{figure}

In Figs.\,\ref{nummm} and \ref{nummassmm} in Sect. 3 we report the number of MPGs as a function of their $\Sigma$  and $\delta$ for the mass-matched samples. In this appendix we show the results for the original samples. In Fig. \ref{numnew} filled points show the numbers of low- and high-$\Sigma$ MPGs in the four $\delta$ bins, corrected for the TSR, SSR, and CSR incompleteness factors. Open magenta and cyan points show the same, but without the completeness correction factors.

\begin{figure}
\includegraphics[width=9.3cm]{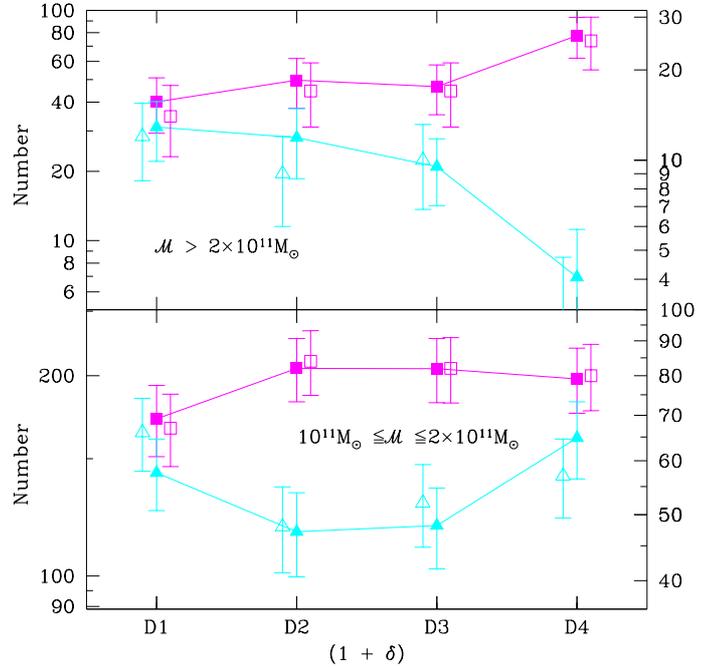}
        \caption{Same as Fig.\ref{numnew}, but for the two bins of stellar mass.}
    \label{nummass}
\end{figure}

In Fig.\,\ref{nummass} we show the number of low- and high-$\Sigma$ MPGs as a function of $\delta$ as in Fig. \ref{numnew}, but for the two subsamples of MPGs with 10$^{11}$M$_{\odot}$ $\leqslant \cal{M} \leqslant$ 2$\times$10$^{11}$M$_{\odot}$ (bottom panel) and with $\cal{M} >$ 2$\times$10$^{11}$M$_{\odot}$ (upper panel).
Overall, the trends we found is similar to the trends derived for the mass-matched samples. 

\section{Negligible effect of the density field definition on our conclusions. }
\begin{figure}
\includegraphics[width=9.5cm]{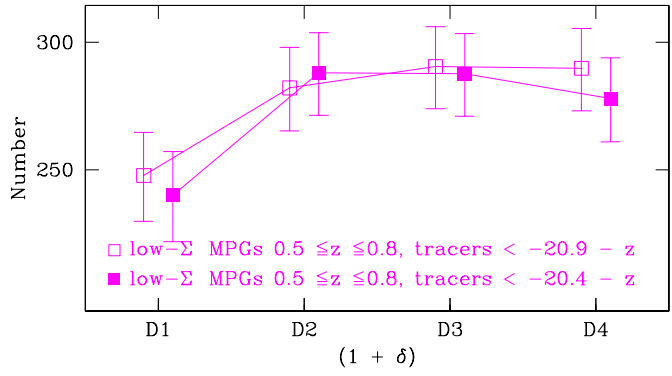}
        \caption{
        Number of low-$\Sigma$ MPGs as a function of $\delta$ in the case the density field is estimated using galaxies with M$_{B} <$ $-20.4 - z$ as tracers (filled magenta squares) and when galaxies with M$_{B} <$ $-20.9 - z$ were used.
}
    \label{conf}
\end{figure}

In this section we show that the choice of the density fields does not affect our conclusions. In Fig.\,\ref{conf} the filled magenta squares  show the number density of low-$\Sigma$ MPGs with 10$^{11}$M$_{\odot}$ $\leqslant \cal{M} \leqslant$ 2$\times$10$^{11}$M$_{\odot}$ as a function of $\delta$, where $\delta$ is computed using as tracer galaxies with M$_{B} \leqslant$ (-20.4 - $z$) (i.e., this is the same as in Fig.\,\ref{numdensf}). Open magenta squares show the same, but for $\delta$ derived using VIPERS galaxies with M$_{B} \leqslant$ (-20.9 - $z$) as tracers. The figure shows that the two functions are in agreement within the errors.
\end{appendix} 
\end{document}